\documentclass[aps,prc,groupedaddress,nofootinbib,twocolumn]{revtex4}

\usepackage[dvips]{graphicx}
\usepackage{amssymb,amsmath,amstext,amsthm,amsfonts}
\usepackage{slashed}

\usepackage[usenames]{color}
\usepackage{pstricks}

\newcommand{\reffig}[1]{Fig.~\ref{#1}}

\newcommand{\refcite}[1]{Ref.~\cite{#1}}

\newcommand{\atom}[2]{\mbox{$^{#1}\text{#2}$}}

\newcommand{\oxygen}{{\atom{16}{O}}}
\newcommand{\plotscale}{.65}

\newcommand{\dd}{\mathrm{d}}

\begin{document}

\title{Neutrino-nucleus interactions in the T2K experiment}

\author{T.~Leitner}
\email{leitner@theo.physik.uni-giessen.de}
\affiliation{Institut f\"ur Theoretische Physik, Universit\"at Giessen, Germany}

\author{U.~Mosel}
\affiliation{Institut f\"ur Theoretische Physik, Universit\"at Giessen, Germany}

\date{June 11, 2010}

\begin{abstract}
  We present a study of neutrino-nucleus interactions at the T2K experiment based on the
  GiBUU transport model. The aim of T2K is to measure $\nu_e$ appearance and
  $\theta_{13}$, but it will also be able to do a precise measurement of $\nu_\mu$
  disappearance. The former requires a good understanding of $\pi^0$ production while the
  latter is closely connected with a good understanding of quasielastic scattering.  For
  both processes we investigate the influence of nuclear effects and particular
  final-state interactions on the expected event rates taking into account the T2K
  detector setup.
\end{abstract}

%\pacs{13.15.+g, 25.30.Pt}

\maketitle

%%%%%%%%%%%%%%%%%%%%%%%%%%%%%%%%%%%%%%%%%%%%%%%%%%%%%%%%%%%%%%%%%%%
%%%%%%%%%%%%%%%%%%%%%%%%%%%%%%%%%%%%%%%%%%%%%%%%%%%%%%%%%%%%%%%%%%%
%%%%%%%%%%%%%%%%%%%%%%%%%%%%%%%%%%%%%%%%%%%%%%%%%%%%%%%%%%%%%%%%%%%

\section{Introduction}

After neutrino oscillations were first observed in atmospheric and solar neutrino rates
\cite{Fukuda:1998mi,Ahmad:2001an}, an extensive experimental program has started aiming at
the precise determination of $\nu$ masses and mixing angles. Of the three mixing angles,
$\theta_{13}$ is still unknown, only upper limits have been placed so far.  Still unknown
is also the sign of one of the two mass squared differences, namely $\Delta m_{32}^2$, and
thus the mass hierarchy (cf., e.g., the review article in \refcite{Amsler:2008zzb}).

The T2K experiment \cite{t2k}, which has just started its operation, is a long-baseline
high-precision neutrino-oscillation experiment. Its major aim is to measure $\nu_e$
appearance and thus $\theta_{13}$ but also to provide a precise measurement of $\nu_\mu$
disappearance to improve $\theta_{23}$ and $\Delta m_{23}^2$. Its target material - common
to all modern detectors - consists of heavy nuclei.  This causes a major difficulty:
Particles produced in neutrino interactions can reinteract before leaving the nucleus and
can be absorbed or change their kinematics or even their charge before being detected. Nuclear
reinteractions limit our ability to identify the reaction channel by changing the
topology of the measured hadronic final state. Consequently, the detected rates on nuclei
are changed significantly compared to the ones on free nucleons.  

Appearance experiments like T2K search for a specific neutrino flavor in a neutrino beam
of different flavor. The flavor of the neutrino can only be determined from the charged
lepton it produces in the interaction.  $\pi^0$ production events in neutral current (NC)
reactions are a source of background in $\nu_e$ appearance searches in a $\nu_\mu$ beam
because they might be misidentified as charge current $(\nu_e,e^-)$ interactions. A
precise and well-tested model for NC$\pi^0$ production on nuclei is thus necessary.

The oscillation probability depends directly on the neutrino energy: $\nu_\mu$
disappearance experiments search for a distortion in the neutrino flux in the detector
positioned far away from the source. From the flux difference, one gains information about
the oscillation probability and with that about $\theta_{23}$ and $\Delta m_{23}^2$.
However, the neutrino energy cannot be measured directly but has to be reconstructed from
the final-state particles that are detected.  Usually, LBL experiments use charged current
quasielastic (CCQE) scattering events for the energy reconstruction. Their identification
is also influenced by nuclear effects which have to be understood and considered in the
experimental analysis. Thus, to extract the oscillation parameters from the measured
particle yields, the experimental analyses have to rely on models for the neutrino-nucleus
interaction (see, e.g., the proceedings of the NUINT conference \cite{nuint09_proc}).

In the T2K experiment, JPARC's high intensity $\nu_{\mu}$ beam is directed toward the
Super-Kamiokande water Cherenkov detector which is located about 300 km away. In addition,
T2K has a high-resolution near detector complex to determine the energy spectrum, the
flavor content of the beam and also neutrino cross sections in particular by measuring
NC$\pi^0$ and CCQE on oxygen. The energy of the neutrino beam extends up to a few GeV with
a maximum at about 600 MeV.

In T2K's case, the near and the far detectors are conceptually very different (water
Cherenkov vs.\ tracking detector) which implies very different event selection schemes.
Thus, a quantitative understanding of neutrino-nucleus cross sections is mandatory.

While the Giessen Boltzmann-Uehling-Uhlenbeck (GiBUU) transport model has already been widely applied to general questions of
neutrino interactions with nuclei \cite{Leitner:2006ww,Leitner:2008ue,leitner_phd} and has
also looked in some more detail at present experiments such as MiniBooNE and K2K
\cite{Leitner:2008wx,Leitner:2010kp} in this paper we primarily investigate the
implications of the different detector techniques on measured NC$\pi^0$ and CCQE spectra
at T2K taking into account in-medium modifications and final-state interactions.

%%%%%%%%%%%%%%%%%%%%%%%%%%%%%%%%%%%%%%%%%%%%%%%%%%%%%%%%%%%%%%%%%%%
%%%%%%%%%%%%%%%%%%%%%%%%%%%%%%%%%%%%%%%%%%%%%%%%%%%%%%%%%%%%%%%%%%%

\section{Neutrino-nucleus scattering in the GiBUU model}

In-medium modifications and final-state interactions inside the target nucleus are known
to be the major source of systematic errors in LBL neutrino experiments
\cite{nuint09_proc}.  Thus, the theoretical understanding of nuclear effects is essential
for the interpretation of the data and only by applying state-of-the-art models it is
possible to minimize the systematic uncertainties in neutrino fluxes, backgrounds and
detector responses.

For this aim, we apply the GiBUU transport model which is a unified framework for the
description of a wide range of nuclear reactions from heavy-ion collisions to pion and
electron/photon scattering and has been widely tested and validated
\cite{Teis:1996kx,Lehr:1999zr,Falter:2004uc,Buss:2006vh} (see \refcite{gibuu} for more
applications). Within this approach, we treat neutrino-nucleus scattering as a two-step
process. In the initial-state step, the neutrinos interact with nucleons embedded in the
nuclear medium.  In the final-state step, the produced particles are propagated through
the nucleus undergoing complex final-state interactions (FSI). Here, we present only a
short review of our model and refer to \refcite{Leitner:2008ue,gibuu} for details.

\begin{figure}[tb]
  \includegraphics[scale=\plotscale]{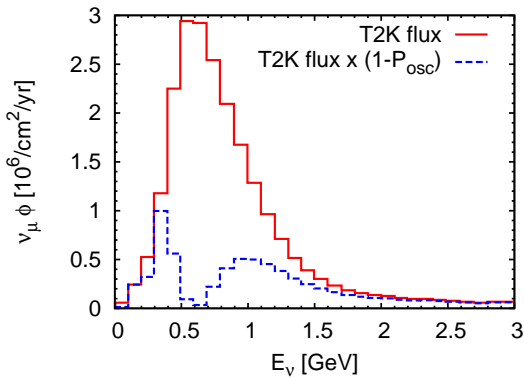}
  \caption{(Color online) Neutrino energy flux for the T2K experiment. The solid line taken from Fig.\ 6 (b)
    of \refcite{Itow:2001ee} shows the unoscillated flux while the dashed line shows the oscillated flux, i.e., the flux multiplied by $(1-P_\text{osc})$ (see text for details).
    \label{fig:neutrino_flux}}
\end{figure}
The energy distribution of the T2K beam is shown in \reffig{fig:neutrino_flux} by the
solid line. The dashed line shows the effect of neutrino oscillations on the flux: The
probability that $\nu_\mu$ remains as $\nu_\mu$ is
\begin{equation}
  P(\nu_\mu \to \nu_\mu) = 1 - P_\text{osc} = 1 - \sin^2 2 \theta_{23} \sin^2 \left( \frac{\Delta m_{23}^2 L}{4 E_\nu} \right),
\end{equation}
using the atmospheric mixing parameters $\theta_{23} = 45^\circ$, $\Delta m_{23}^2=2.5
\times 10^{-3}$ eV$^2$, and a distance of $L=295$ km.

\begin{figure}[tb]
  \includegraphics[scale=\plotscale]{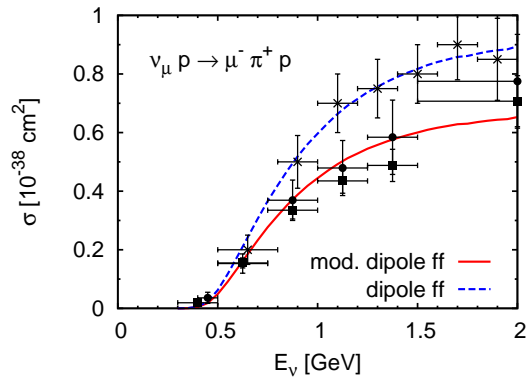}
  \caption{(Color online) Total $\nu_\mu p \to \mu^- \pi^+ p$ cross section as a function
    of the neutrino energy compared to the pion production data of of ANL
    (Refs.~\cite{Barish:1978pj} ($\bullet$), \cite{Radecky:1981fn} ($\blacksquare$)) and
    BNL (\cite{Kitagaki:1986ct} ($\times$)). The solid line has been obtained with a form
    factor fitted to the ANL data, the dashed line is fitted to the BNL data.}
  \label{fig:Delta_sigmatot}
\end{figure}

In this energy region, the elementary $\nu N$ reaction is dominated by two processes:
quasielastic scattering and the excitation of the $\Delta$ resonance (P$_{33}$(1232)). The
vector form factors are well constrained by electron scattering data (see
\refcite{Leitner:2008ue} for details) while we have a significant uncertainty in the axial
form factors: Goldberger-Treiman relations have been derived for the axial couplings, but
they do not give information about the $Q^2$ dependence. The $Q^2$ dependence of the
$\Delta$ axial form factor is fitted to either ANL or BNL bubble chamber
neutrino-scattering $\dd \sigma/\dd Q^2$ data for the $\nu_\mu p \to \mu^- \pi^+ p$
reaction. Fig.~\ref{fig:Delta_sigmatot} shows the integrated cross section together with
the data.  Note that the solid curve fits the ANL data (modified dipole form factor),
while the dashed curve fits the BNL data (dipole form factor). Thus, the latter would
obviously lead to higher pion production cross section also on the nucleus. We refer the
reader to Chapter 5 of \refcite{leitner_phd} for an extended discussion. Recently, a
combined form factor fit to both ANL and BNL data has also been performed in
\refcite{Graczyk:2009qm,Hernandez:2010bx}.

In the nuclear medium, the neutrino-nucleon cross sections are modified. Bound nucleons
are treated within a local Thomas-Fermi approximation. They are exposed to a mean-field
potential that depends on density and momentum. We also take Pauli blocking and the
medium-modified spectral functions of the outgoing hadrons into account. Our model for
neutrino-(bound)nucleon scattering is described in detail in \refcite{Leitner:2008ue}.

After the initial neutrino-nucleon interaction, the produced particles propagate through
and out of the nucleus. During propagation they undergo complex FSI simulated with the
coupled-channel GiBUU transport model (for details, see \refcite{gibuu} and references
given there).  It models the space-time evolution of a many-particle system in a
mean-field potential including elastic and inelastic collisions between the particles, and
also particle decays into other hadrons.

To summarize, the GiBUU model provides a theory-based, consistent and well-tested
treatment of in-medium modifications and final-state interactions. It successfully
describes electron-and photon-induced reactions over a wide energy range. Because these
reactions are quite similar to the $\nu A \to X$ reaction in that the incoming particle
interacts with all target nucleons and the vector couplings are the same, we consider this
as an important benchmark for the description of neutrino-induced reactions.

%%%%%%%%%%%%%%%%%%%%%%%%%%%%%%%%%%%%%%%%%%%%%%%%%%%%%%%%%%%%%%%%%%%%%%%%%%%%%%%%%%%%%%%%%%

\section{NC$1\pi^0$}

\begin{figure}[tbp]
  \centering
  \includegraphics[width=\columnwidth]{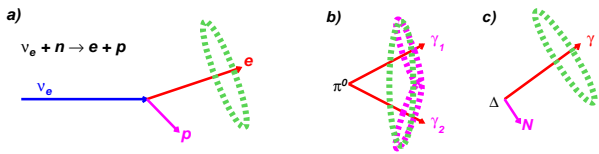}
  \caption[]{(Color online) Panel (a) shows the ``$\nu_e$ appearance signal'' event in a Cherenkov detector
    while panels (b) and (c) are possible backgrounds. Taken from
    \refcite{Djurcic:2006sf}.\label{fig:Cherenkov_nue_appea_background}}
\end{figure}

The main task in a $\nu_e$ appearance experiment as T2K is to detect electron neutrinos in
a (almost) pure $\nu_\mu$ beam in the Super-Kamiokande detector. The signal event is the
$\nu_e$ CCQE interaction as indicated in \reffig{fig:Cherenkov_nue_appea_background} (a).
However, not all of the detected $\nu_e$ events have their origin in neutrino oscillations
but are already present in the initial neutrino beam, e.g., through muon and kaon decays.
A good understanding of the neutrino beam composition is therefore essential for $\nu_e$
appearance searches.

Even more significant at low reconstructed $E_\nu$ are misidentified events, mainly
because a Cherenkov detector like Super-Kamiokande cannot distinguish between a photon and
an electron.  Thus, $\nu_\mu$ induced NC $\pi^0$ production,\footnote{We
  recall that in NC events only the outgoing hadrons and/or their decay products are seen,
  but not the outgoing neutrino.} where the $\pi^0$ decays into two $\gamma$s, is the
major source of background when one of the photons is not seen or both Cherenkov rings
overlap.  This process is illustrated in \reffig{fig:Cherenkov_nue_appea_background} (b).
Additional background comes from the excitation of a $\Delta$ resonance via a NC interaction followed by its radiative decay, $\Delta \to \gamma N$, which also
leads to a final state with a photon (see \reffig{fig:Cherenkov_nue_appea_background}
(c)).\footnote{See \refcite{Leitner:2008fg} for a brief discussion of the radiative
  $\Delta$ decay.}

A proper understanding of neutrino-induced pion production, especially NC$\pi^0$, is
therefore essential in $\nu_e$ appearance experiments.  T2K requires the NC$\pi^0$ cross
section to be known to 10\% accuracy for the resulting error on the oscillation parameters
to be comparable to that from statistical uncertainties \cite{AguilarArevalo:2006se}. This
has triggered a lot of experimental activity toward direct NC$\pi^0$ cross section
measurements in the last years to be used as direct input for the oscillation analysis
\cite{Nakayama:2004dp,AguilarArevalo:2008xs,Kurimoto:2009wq,Anderson:2009zz}.

\begin{figure}[tb]
  \includegraphics[scale=\plotscale]{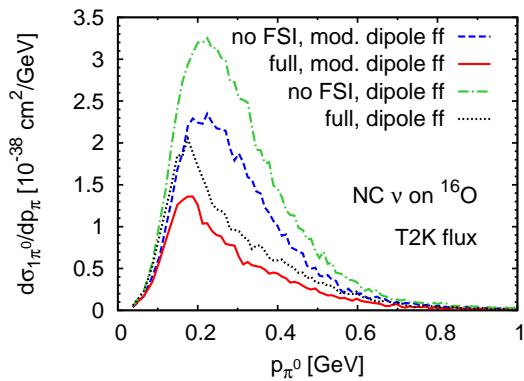}
  \caption{(Color online) NC induced single-$\pi^0$ production on \oxygen{} as a function of the pion
    momentum averaged over the T2K flux. The dashed and the solid lines denote the
    calculation with a modified dipole form factor (fitted to ANL data) for the $\Delta$
    resonance. The dash-dotted and dotted lines are obtained with a
    dipole form for the form factor (fitted to the BNL data).  
    \label{fig:T2K_NC_ff}}
\end{figure}

In the following we study the influence of nuclear effects and final-state interactions on
NC$\pi^0$ production (see also \refcite{Leitner:2006sp}).  \reffig{fig:T2K_NC_ff} shows
the results for NC single-$\pi^0$ production off \oxygen{} averaged over the incoming,
unoscillated T2K energy distribution \cite{Itow:2001ee}.  Comparing the dashed with the
solid line (results without FSI and spectral function vs.~full calculation), one finds a
significant difference.  The shape is caused by the energy dependence of the pion
absorption and rescattering cross sections.  Pions are mainly absorbed via the $\Delta$
resonance, i.e, through $\pi N \to \Delta$ followed by $\Delta N \to NN$. This explains
the reduction in the region around $p_\pi=0.2-0.5$ GeV.  Pion elastic scattering $\pi N
\to \pi N$ reshuffles the pions to lower momenta and leads also to charge exchange
scattering into the charged pion channels. Furthermore, we show in this plot the result
for our second form factor set, namely the dipole form factor (dash-dotted and dotted
lines). The difference between the two sets reflects the uncertainty caused by the
discrepancy in the old ANL and BNL data.

The total flux averaged NC$1\pi^0$ cross section is $(0.34 - 0.48)\,\cdot\,10^{-38}\,
\text{cm}^2$ where the range again gives the uncertainty from the axial form factor.

The vast majority of the pions come from initial $\Delta$ excitation, pion production
through higher resonances plays almost no role. $\pi^0$ can be produced in our approach
not only through resonances but also through initial NC elastic scattering, i.e.,
final-state nucleons can rescatter in the nucleus and produce pions; this accounts for
$\approx$ 2\%. Missing in our approach is deep-inelastic scattering and any
non-resonant single-$\pi$/double-$\pi$ etc.\ background. However, we do not expect a
significant contribution of those at the rather low T2K energy. Coherent pion production
is also possible, but its contribution is only of the order of
$(0.02-0.04)\,\cdot\,10^{-38}\,\text{cm}^2$ \cite{hernandezNUINT09Talk}. Thus, our model
accounts for the relevant contributions.

\begin{figure}[tb]
  \includegraphics[scale=\plotscale]{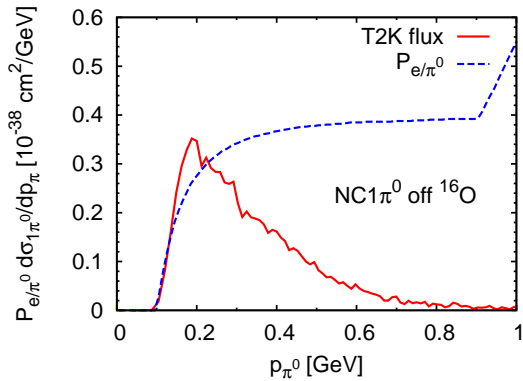}
  \caption{(Color online) NC induced single-$\pi^0$ production cross sections on \oxygen{}
    averaged over the T2K flux and multiplied by the misidentification
    probability (dashed line, taken from \refcite{Okamura:2010zz}) as a function of the
    pion momentum.
    \label{fig:T2K_NC_misid}}
\end{figure}

A crucial problem in any such experiment with Cherenkov counters is that the decay photons
of neutral pions can be misidentified as electrons. We therefore, investigate in
\reffig{fig:T2K_NC_misid} the probability that a misidentified $\pi^0$ is counted as a
$\nu_e$ appearance event.  The probability that a $\pi^0$ cannot be distinguished from
$e^\pm$ is given by the dashed line (taken from Fig.~2 of \refcite{Okamura:2010zz}).  It
mainly depends on the opening angle of the two decay photons. The solid line shows the
weighted cross section averaged over the T2K flux and calculated using the elementary ANL
data as input. The total cross section for misidentified events is now
$0.09\,\cdot\,10^{-38}\, \text{cm}^2$ and thus 26\% of the true pion events.

%%%%%%%%%%%%%%%%%%%%%%%%%%%%%%%%%%%%%%%%%%%%%%%%%%%%%%%%%%%%%%%%%%%%%%%%%%%%%%%%%%%%%%%%%%

\section{CCQE}

The other major aim of the T2K experiment is the precise measurement of $\nu_\mu$
disappearance and the corresponding mixing parameters. This task requires an equally
precise measurement of the neutrino flux both at the near and the far detector. From the
flux difference, one can extract the oscillation parameters. The neutrino energy is
reconstructed using the CCQE reaction, thus, the related challenge is to identify
\emph{true} CCQE events in the detector, i.e., muons originating from an initial QE
process $\nu_\mu n \to \mu^- p$.  The difficulty comes from the fact that the true CCQE
events are masked by FSI in a detector built from nuclei. In T2K the QE event selection is
different in the near and far detector: in general, the near detector counts all events as
CCQE-like that have a single proton track but no pions. In the water Cherenkov detector
Super-Kamiokande, serving as far detector, CCQE-like means, that no pion is seen in the
event.  In both detectors the FSI lead to misidentified events, e.g., an initial $\Delta$
whose decay pion is absorbed or undergoes ``pion-less decay'' looks ``CCQE-like''.

\begin{figure}
  \centering
  \includegraphics[scale=\plotscale]{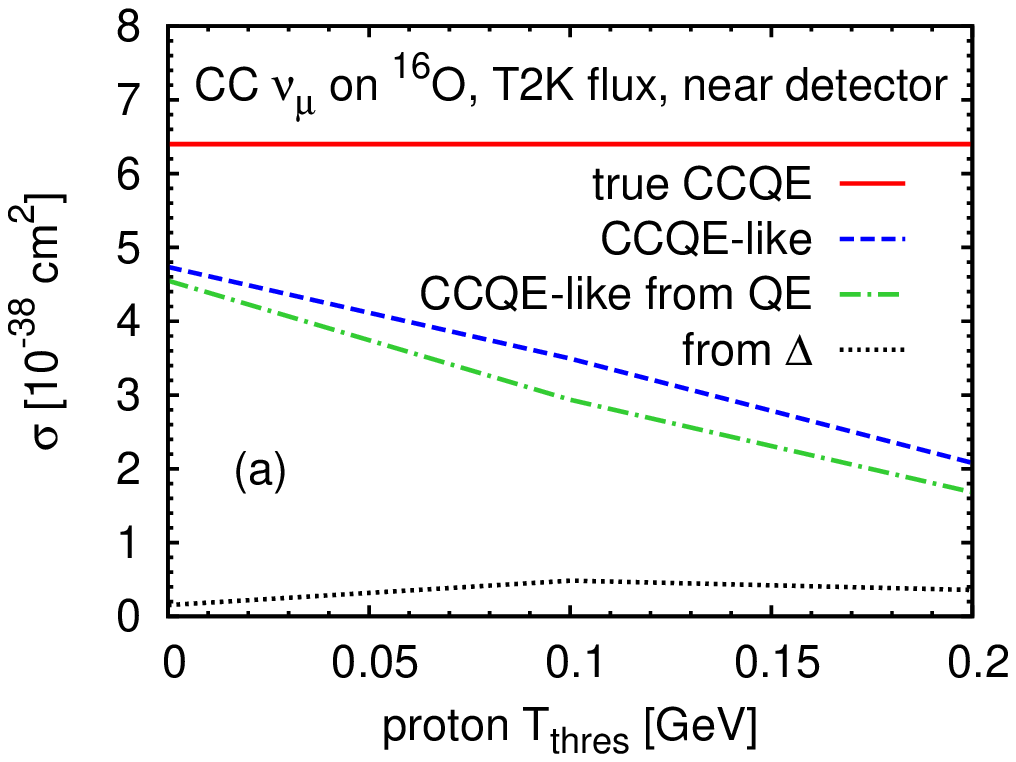}
  \includegraphics[scale=\plotscale]{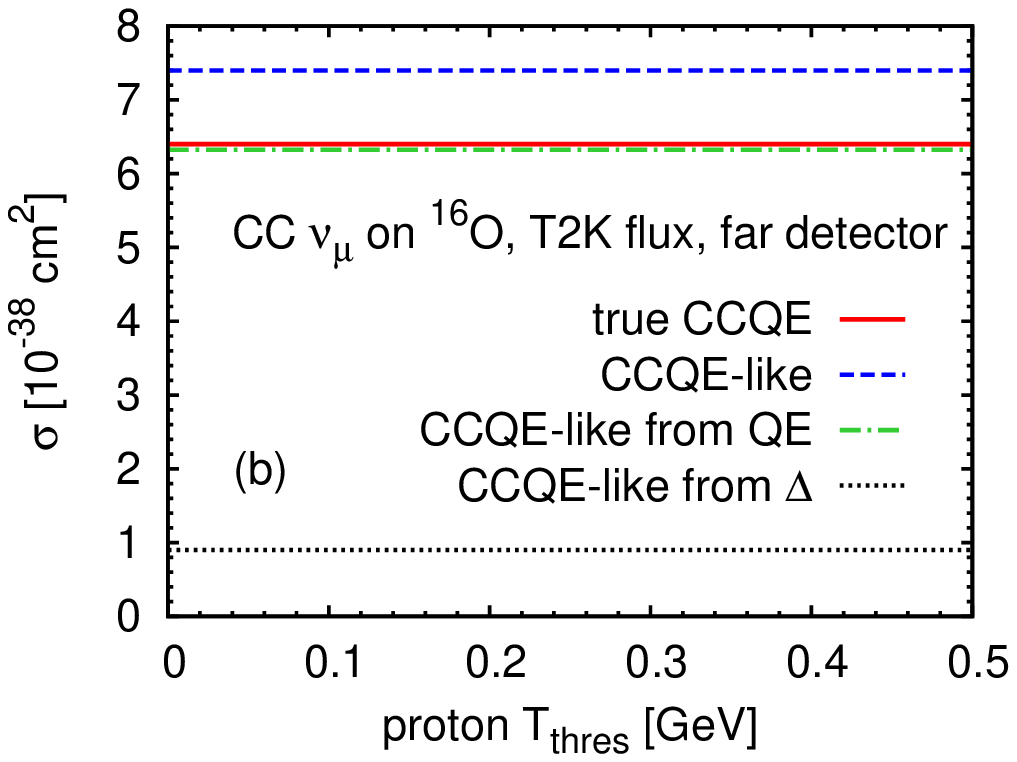}
  \includegraphics[scale=\plotscale]{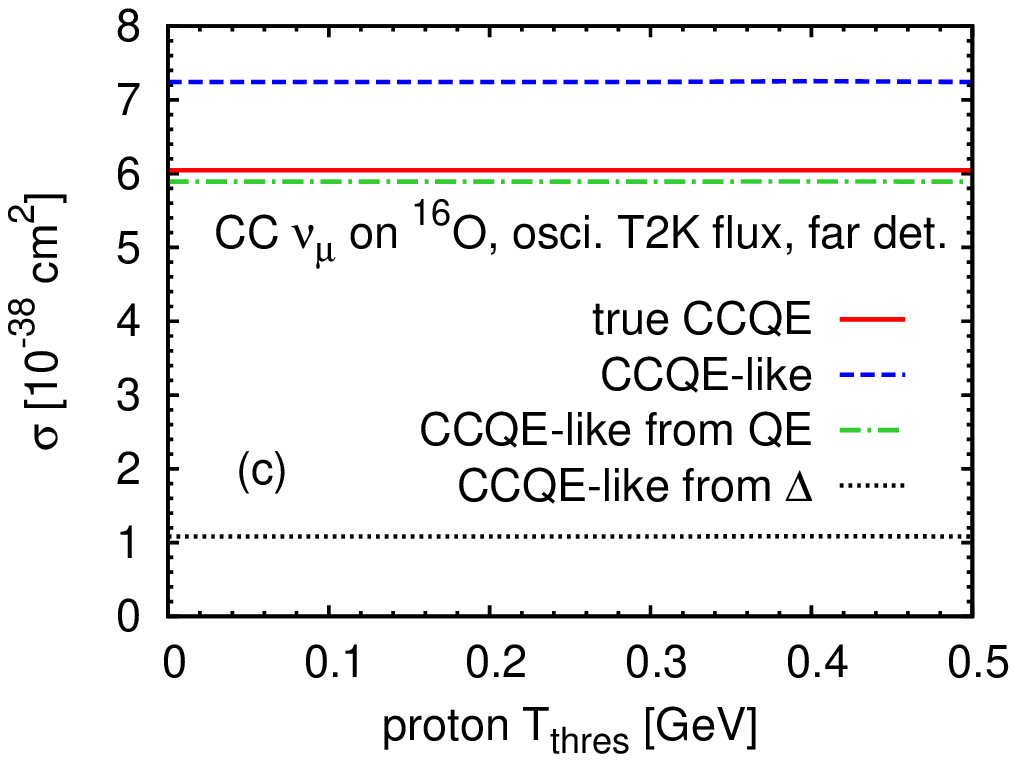}
  \caption{(Color online) Total flux averaged CCQE cross section on \oxygen{} (solid
    lines) versus the proton kinetic energy detection threshold. Shown are besides the
    true CCQE cross section two different methods on how to identify CCQE-like events in
    experiments (dashed lines).  Panel (a) shows the method applied in the near detector
    (tracking detector), panel (b) the one applied in the far detector (Cherenkov detector). Panel (c)
    shows the result for the far detector but averaged over the oscillated flux shown in
    \reffig{fig:neutrino_flux}.  The contributions to the CCQE-like events are also
    classified (CCQE-like from initial QE (dash-dotted lines), from initial $\Delta$ (dotted
    lines)).
    \label{fig:QElike_protonThres}}
\end{figure}

\begin{figure}
  \centering
  \includegraphics[scale=\plotscale]{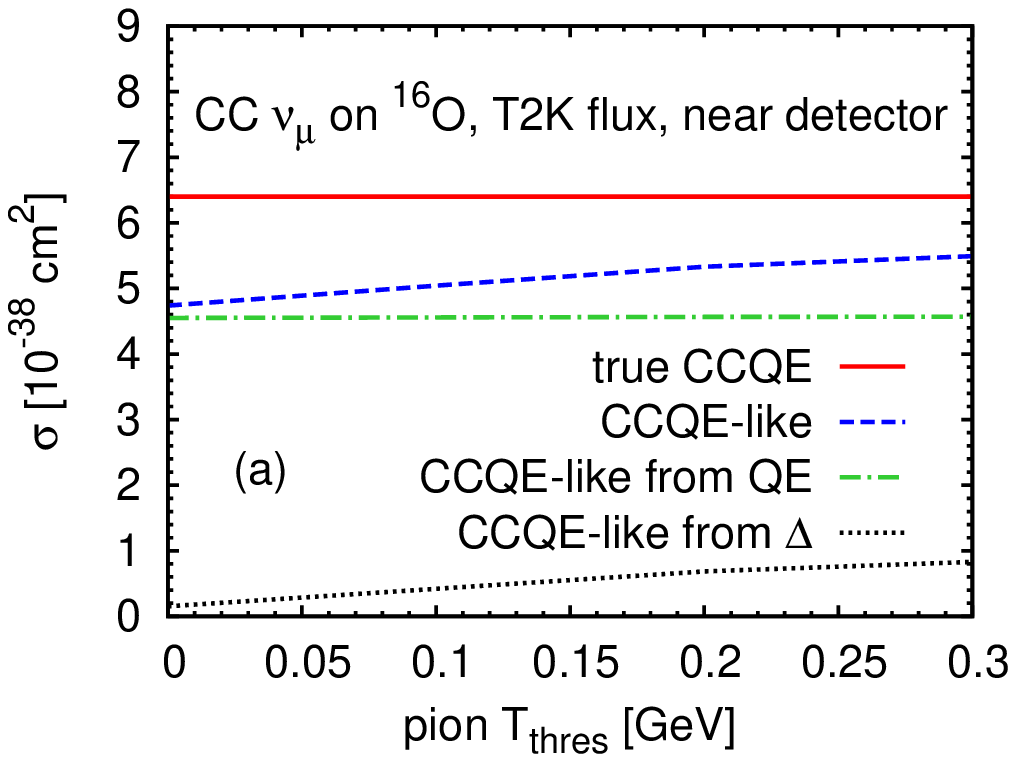}
  \includegraphics[scale=\plotscale]{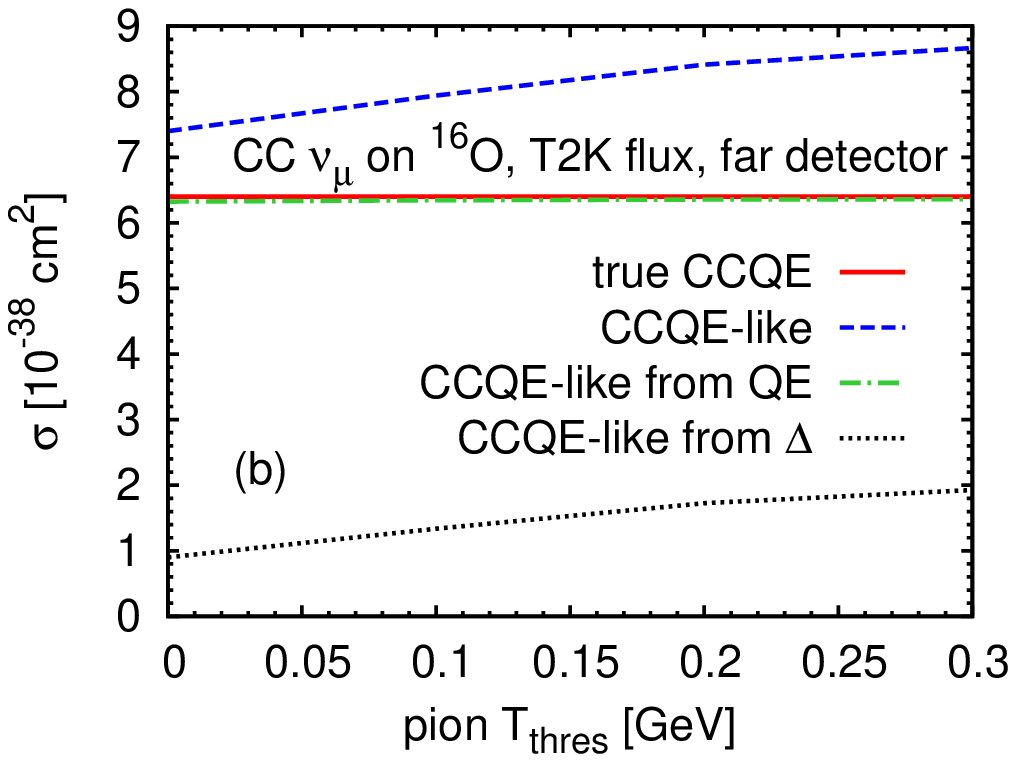}
  \includegraphics[scale=\plotscale]{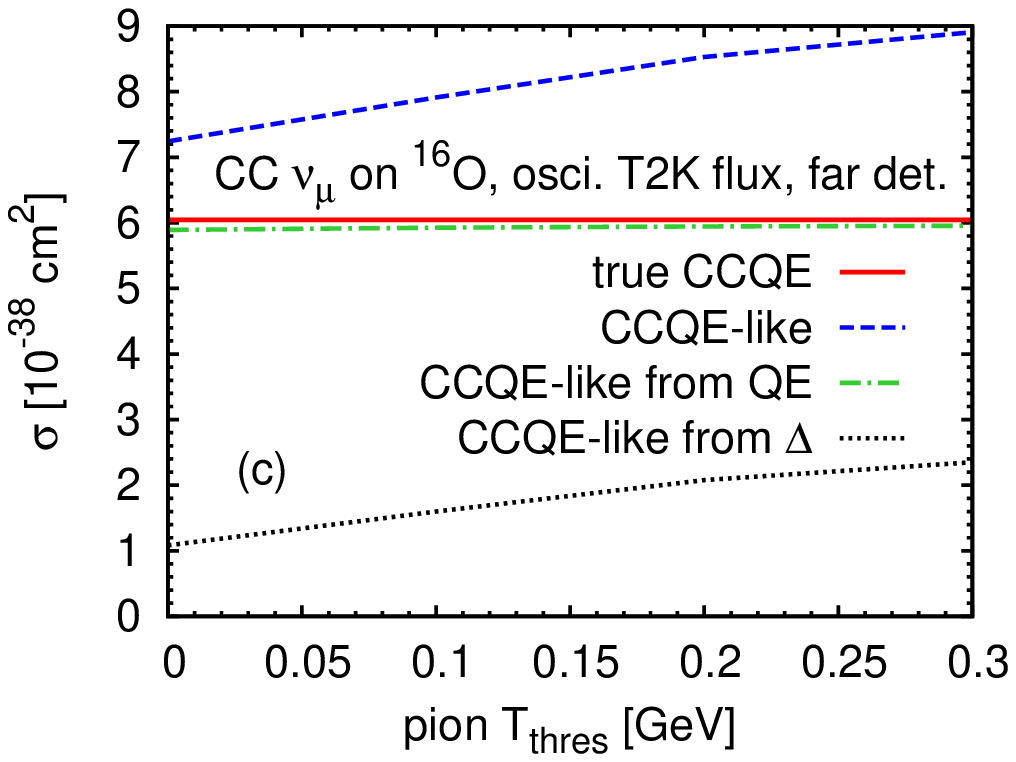}
  \caption{(Color online) Total flux averaged CCQE cross section on \oxygen{} (solid lines)
    versus the charged pion kinetic energy detection threshold. The lines are as in
    \reffig{fig:QElike_protonThres}.
    \label{fig:QElike_chPionThres}}
\end{figure}

In the following, we investigate the quality of these event selection methods considering
also pion and proton kinetic energy detection thresholds.  \reffig{fig:QElike_protonThres}
(a) shows the event selection for the near detector (tracking detector: CCQE-like = proton
but no pion) while the far detector is addressed in panel (b) (Cherenkov detector:
CCQE-like = no pion). The flux averaged cross sections are plotted versus the proton
kinetic energy detection threshold.  The ``true CCQE'' events are denoted with the solid
lines, the CCQE-like events by the dashed ones. In addition, those are separated in the
two main contributions, namely CCQE-like from initial QE scattering (dash-dotted lines)
and CCQE-like from initial $\Delta$ excitation (dotted lines).

In the near detector considerably fewer CCQE-like than true CCQE events are detected (panel
(a), the difference between the dashed and the solid line). The final-state interactions of the
initial proton lead to secondary protons, or, via charge exchange to neutrons which are
then not detected as CCQE-like any more (trigger on \emph{single} proton track).  However,
the amount of fake events in the CCQE-like sample is small (the dashed and dash-dotted lines
are close). Thus, the observation of the proton helps to restrict the background, but also
leads to an underestimate of the true CCQE cross section by at least 25\%. This difference
has to be reconstructed with event generators that have to be very realistic in
describing the in-medium nucleon-nucleon interactions. The results are obviously very
sensitive to the proton kinetic energy detection threshold because the event selection
procedure explicitly triggers on a single proton. At a realistic detection threshold of
0.2 GeV proton kinetic energy only about 30\% of the total QE cross section is
measured and 70\% has to be reconstructed.

The far detector is able to detect almost all true CCQE (panel (b), solid vs.\ dash-dotted
lines agree approximately) but sees also a considerable amount of ``fake CCQE'' (or
``non-CCQE'') events (panel (b), the dashed line is roughly 15\% higher than the solid
line). These fake events have to be removed from the measured event rates by means of
event generators, if one is interested only in the true QE events. It is obvious that this
removal is the better the more realistic the generator is in handling the in-medium
$\pi-N-\Delta$ dynamics. For this detector the results do not depend on the proton kinetic
energy threshold because the event selection is independent of the proton.

\reffig{fig:QElike_protonThres} (c) shows a similar scenario as (b) but now averaged over
the oscillated T2K flux.  We find almost no difference of the total CCQE-like cross
section compared to panel (b), however, the contributions are slightly shifted (more
$\Delta$, less true QE). This is caused by the two-bump structure of the oscillated flux
shown in \reffig{fig:neutrino_flux}: The lower energy bump leads mainly to true CCQE
events but --- because of the lower energy --- with reduced cross section (solid line). However, the pion cross section and with that the CCQE-fake cross section is
increased owing to the higher energy bump (dotted line).  However, overall, the effect of
neutrino oscillations on the CCQE cross section is only minor.

\reffig{fig:QElike_chPionThres} shows the same cross sections as the previous plot now
versus the pion kinetic energy detection thres\-hold. Both detector types require 'no pions'
for an event to be CCQE-like, thus, the CCQE-like cross section increases with increasing
threshold because then more and more pions are not seen in the detector.  More exactly,
the $\Delta$ contribution to the CCQE-like cross section (dotted lines), i.e., the fake
events, while the QE contribution (dash-dotted lines) stays constant. Also here the
average over the oscillated flux (panel (c)) is very similar to the one over the
unoscillated flux (panel (b)), and again the single contributions which add up to the
dashed line are slightly different.

We note that the uncertainty from the $\Delta$ axial form factor is only of minor
importance here because the results are dominated by QE scattering.

Finally, we discuss the implications for the neutrino energy reconstruction which is of
major importance for $\nu_\mu$ disappearance measurements.  The neutrino energy is
reconstructed from the CCQE-like events. The preceding discussion and also our detailed study
in \refcite{Leitner:2010kp} have shown that that the QE identification is worse in
Cherenkov detectors than in tracking detectors: The CCQE-like sample in the T2K far
detector contains a significant amount of non-QE induced events while the CCQE-like sample
in the T2K near detector is rather clean but significantly reduced. The neutrino energy is
reconstructed from the measured muon properties of the full CCQE-like sample assuming
quasifree kinematics. The non-QE induced fraction of the CCQE-like sample causes a shift
of the reconstructed energy to values lower than the true energy \cite{Leitner:2010kp}.
This effect is relevant in the far detector. Because of the conceptional difference
between the two detectors, one reconstructs different energy spectra in the near and in
the far detector with a larger systematic error in the latter.  In
\refcite{Leitner:2010kp} we have shown that at a neutrino energy of 0.7 GeV, corresponding
roughly to the flux-maximum in the T2K experiment, the standard deviation for the
reconstructed energy from the true energy is 18\% for the tracking (near) detector,
whereas it is 23\% for the Cherenkov (far) detector. Because for an extraction of the
oscillation parameters the flux has to be compared at the same energy, the overall
combined standard deviation of the energy is thus $\sqrt{18^2 + 23^2}\,$\% = 29\%. This
inaccuracy enters directly into the determination of the neutrino oscillation parameters
which underlines the need for a quantitative understanding of the described effects.

\section{Conclusions}

In this work we have applied the GiBUU hadronic transport model to the T2K experiment. For
their physics goals, T2K needs a good description of both the NC$\pi^0$ and the CCQE cross
section. The accuracy of the former depends strongly on the choice for the $\Delta$ axial
form factor. The NC$\pi^0$ spectra are strongly modified by FSI so that a good
understanding of these is also mandatory in the experimental analysis. Furthermore, our
analysis shows that about 26\% of all pion events contribute to misidentified electron
appearance events.

We have furthermore shown that the measured CCQE cross section depends strongly on the
detector setup and that it is closely entangled with the CC$1\pi^+$ cross section. This
has consequences for the neutrino energy reconstruction in T2K where near and far detectors
are conceptually different: The QE identification is worse in Cherenkov detectors than in
tracking detectors and, therefore, the reconstructed energy has a wider error in the far
than in the near detector. We find that an error of about 29\% is to be expected for the
reconstructed energy in the oscillation parameter extraction.

To conclude, in-medium effects and, in particular, final-state interactions influence the
measured rates significantly. Their understanding --- within a consistent and well-tested
model --- is crucial for the data analysis and interpretation of the T2K experiment.

\begin{acknowledgments}
  We thank O.~Lalakulich for fruitful discussions.  This work has been supported by the
  Deutsche Forschungsgemeinschaft.
\end{acknowledgments}

% bibliography

\end{document}